# Electricity generated from Ambient Heat by Pencils

Zihan XU[1#*] and Guo'an Tai[2#]

[1] Carbon Source Technology, LLC, Shenzhen, China
[2] The State Key Laboratory of Mechanics and Control of Mechanical Structures,
    Nanjing University of Aeronautics and Astronautics, Nanjing, China

[*] zihan.xuu@gmail.com
[#] These authors contributed equally to this paper

**Abstract:**

**The idea of generating electricity from ambient heat has significant meanings for both science and engineering. Here, we present an interesting idea of using pencil leads, which are made of graphite and clay, to generate electricity from the thermal motion of ions in aqueous electrolyte solution at room temperature. When two pencil leads were placed in parallel in the solutions, output power of 0.655, 1.023, 1.023 and 1.828 nW were generated in 3 M KCl, NaCl, NiCl$_2$ and CuCl$_2$ solutions, respectively. Besides, we also demonstrate that two pieces of reduced graphene oxide films and /or few-layer graphene films can generate much more electricity when dipped into the solutions, while there was no electrodes contact with the solution. This finding further verified that the electricity was not resulted from the chemical reaction between the electrodes and the solutions. The results also demonstrate that ambient thermal energy can be harvested with low dimensional materials, such as graphene, or with the**

**surface of solid material without the presence of temperature gradient. However, the mechanism is still unclear.**

Energy conversion techniques such as photovoltaic, thermoelectric and piezoelectric are very attractive for science and engineering as they are possible way out the energy and environment crisisthat we are suffering[1]. Recently, we found that electricity can also be generated from the limitless ambient heat by graphene[2] and reduced graphene oxide (RGO)[3], which opened the door for utilizing the thermal energy to meet the need of clean energy. However, there was highly doubt that the electricity generated came from chemical reaction between electrodes and the graphene instead of the ambient heat[4-6]. Herein, we present an interesting idea of using pencils, RGO or graphene to collect energy from the thermal motion of ions in aqueous electrolyte solutions, which are maintained by the energy of the ambient heat, and convert it into electricity. Furthermore, the possibility of chemical reaction was reasonably excluded.

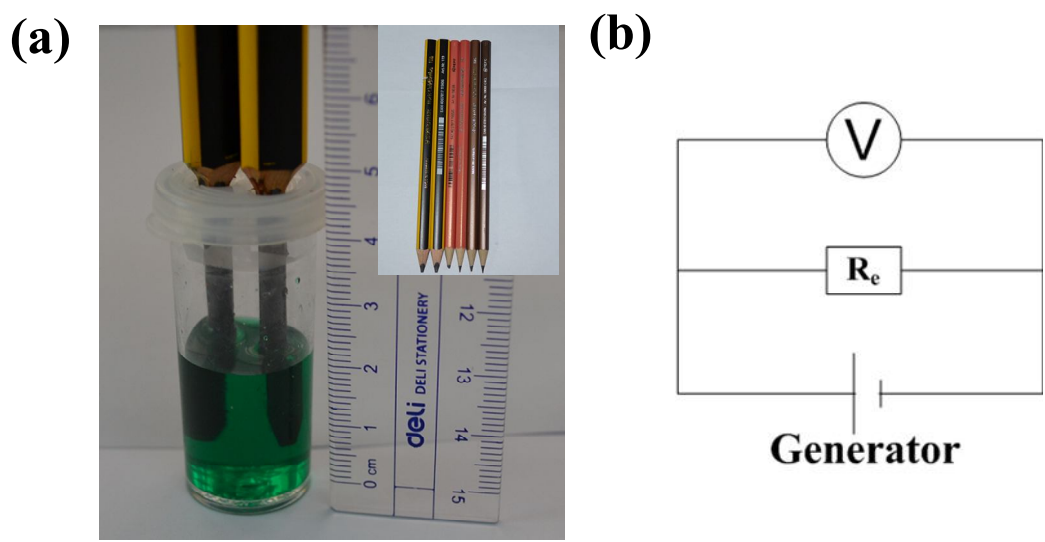

**Fig. 1**. (a) The setup of the experiments and the pencils used in this experiment (insert, from left to right: big HB (STAEDTLER, Art Nr.119, Germany, lead diameter is 0.5 cm), small 2B (STAEDTLER, Art Nr.180-2B, Germany, lead diameter is 0.3 cm), small HB pencil

(STAEDTLER, Art Nr.180HB,-C4CL, Germany, lead diameter is 0.3 cm)). (b) The corresponding equivalent circuit.

The experimental setup is shown in Fig. 1, two HB pencils with leads' diameter of about 0.5 cm were pared to let 4 cm pencil leads in one end exposed to the solutions, while about 1 cm lead in the other end are exposed to air and connected with a 220 kohm resistor and a multimeter. In order to give a comparison between different materials and solutions, the 220 kohm resistor was always loaded to the circuits during all the measurements. The two pencils were fixed together to keep the pencil leads separated from each other. The space between two ends was about 0.5 cm, as shown in Fig. 1a. The photographs of big HB, 2B and small HB pencils were shown in the inset of Fig. 1a. The fixed pencils were settled in a plastic vial, and then the solutions were added into it. The corresponding equivalent circuit was shown in Fig.1b. The voltage signals were collected by a Labview program controlled KEITHLEY 2000 multimeter.

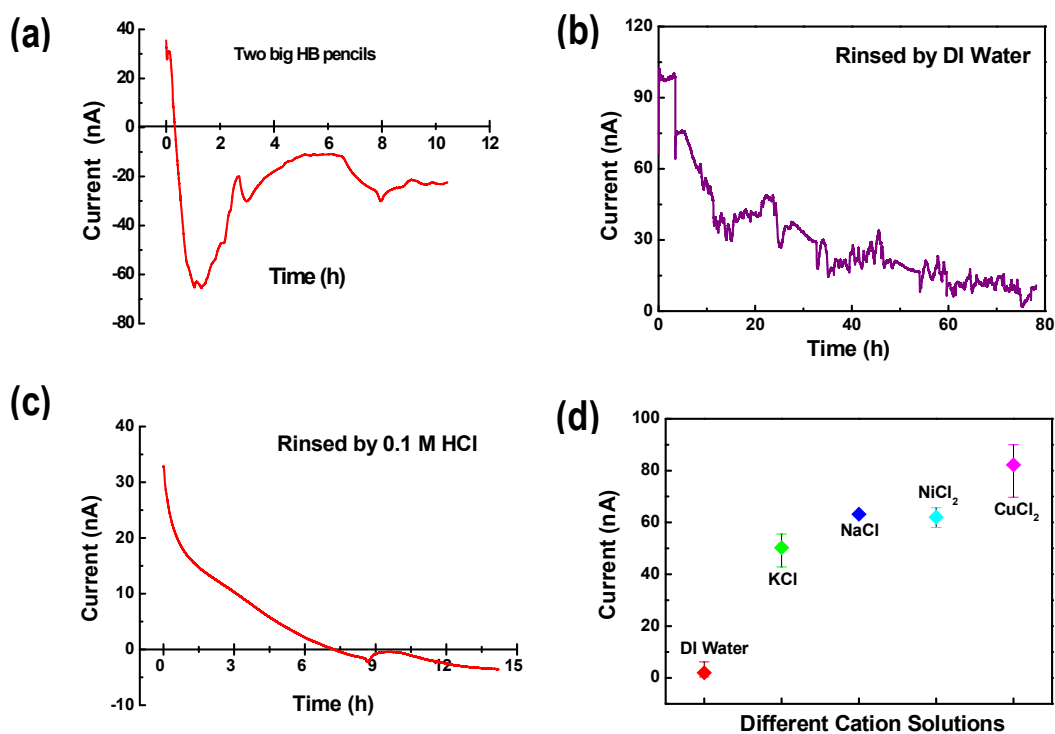

**Fig. 2.** Output of two HB pencils with leads' diameter of around 0.5 cm in parallel immersed into 3 M $CuCl_2$ solution. (a) Output of the pencils. (b) Output of the pencils after the leads were rinsed

by deionized water. (c) Output of the pencils after the leads were rinsed by 0.1 M HCl solution. (d) Output of the pencils in 3 M different cation solutions. The 220 kohm resistor was always loaded to the circuits during all the measurements.

When there was no solution in the vial, the output voltage on the resistor was only tens of microvolt, which can be regarded as noise and suggests that no electricity was generated when no solution was added into the vial. When 4 mL $CuCl_2$ solution (3 M) was poured into the vial, a current of about 40 nA and a power of 0.4 nW were produced from the device. After about half an hour, the current dropped to zero, and then was reversed from positive to negative. Afterward, the current reached a peak up to about -70 nA after the pencil leads were immersed to the solution for about 1.5 h. The reversion reveals the same phenomenon as in the work using same electrodes on the graphene.

When the procedure was performed over 10 h, the output current dropped to about -20 nA. There are two possible reasons for the downtrend: (1) after 10 h, an equilibrium could be established between the symmetrical electrodes; (2) the saturation of absorption of ions on the leads' surface had been established. Then, we rinsed the leads by deionized water to clean the surface of the pencil leads. The output current boomed to about 100 nA (Fig. 2b) and showed the same downtrend as observed in Fig. 2a. The current decreased to about 10 nA after about three days (Fig. 2b). We also rinsed the leads by 0.1 M HCl, but found that the effect was not as good as DI water (Fig. 2c), which might due to that the presence of $Cl^-$ weakened the effect of the washing process.

Different ions were studied to investigate this phenomenon. All the concentration of cations was 3 M and all the measurements were performed at room temperature. All the data was generated at the rate of 1 data/second, and was collected after the pencil leads had been exposed to the solution for 10 minutes. The 220 kohm resistor was loaded to the circuits during all the measurements. The HB-HB pencil pair was firstly dipped to NaCl vial and then NaCl, NiCl and finally $CuCl_2$ vial. The results showed that $Cu^{2+}$ exhibited better ability to produce electricity compared with $K^+$, $Na^+$ and $Ni^{2+}$, which was consistent with the results we got in our previous experiment with graphene[2]. The control experiment with DI water was also performed and a current of about 0.8 nA was obtained which canbe ascribed to a small number of residual cations in the DI water

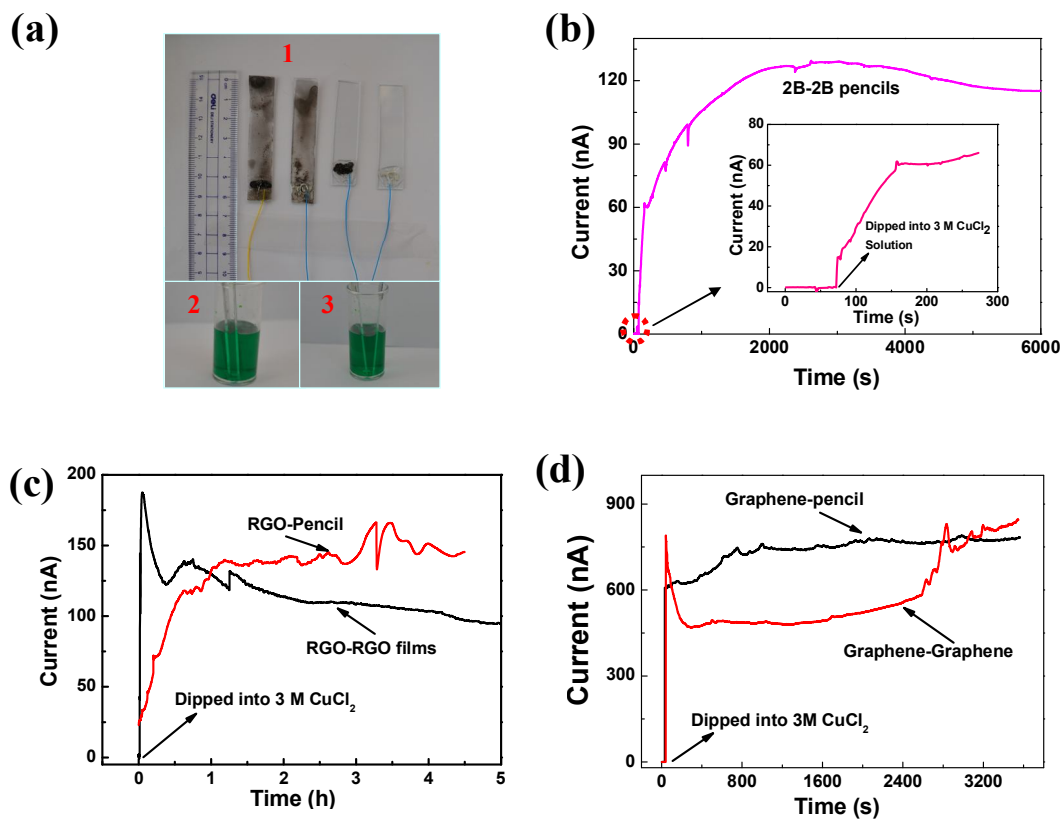

**Fig. 3** (a) Photographs of the samples with graphene and reduced graphene oxide: 1, graphene and RGO film on glass, from left to right: RGO film with graphite as electrode, RGO film with silver as electrode, graphene film with graphite as electrode and graphene with silver as electrode. 2, two RGO films in 3 M $CuCl_2$ solution. 3, two graphene films in 3 M $CuCl_2$ solution; (b) Output of two small 2B pencil leads in 3 M $CuCl_2$ solution; (c) Output of RGO-HB lead pair and RGO-RGO film pair; (d) Output of graphene- HB lead pair and graphene-graphene pair.

Different pencils were also used to investigate theinfluence of the exposed area and the concentration of graphite on generating electricity. When one of the two big HB pencils was replaced by a small HB pencil whose diameter was about 0.3 cm and has the same graphite percentage but smaller exposed area, the induced current will drop a little (Supporting information Fig.S2). Then the two HB pencil leads were replaced by 2B pencils in which graphite concentration is higher than the HB ones. The results in Fig. 3b showed that the induced current was a little higher than that of big HB and small HB pencil. This means that the higher the concentration of graphite, the more the electricity will be generated from these experiments.

We also replaced the pencils with reduced graphene oxide (RGO) and graphene films to verify the finding with pencils. Graphene oxide (GO) was prepared with the improved Hummers Methods and was dispersed in 50% DI water/ 50% ethanol solution. Several droplets were dropped onto the glass. The graphene oxide films were reduced at 250 $^{o}$C in the air [7]. Graphene films (XFnano, Nanjing) were transferred to glass, as shown in Fig. 3a. Silver coated copper wires were fixed to one end of the samples by graphite glue or silver pastes. The pastes and wires were covered with clear paste to avoid the contact with the solution, as shown in Fig. 3a.

When a pencil lead was replaced by RGO film, the current increased to 150 nA sharply (Fig. 3c), which was about 3 times higher than that with pencil-pencil samples. And the current direction was fixed due to the introduction of the asymmetrical structure. When two pencils were replaced by two RGO films with graphite and silver as electrodes respectively, the induced current increased to about 200 nA immediately when they were dipped into the 3 M $CuCl_2$ solution. But the value dropped by 50% after 5 h. The downtrend shows that the asymmetrical structure was not an ideal one as both the films were made of the same material. The graphene samples exhibited the same tendency as RGO samples, but the current were much higher than that from RGO samples, and could reach a value of about 800 nA, which was more than 10 times higher than the value of pencils. We also carried out the experiment with graphene-graphene pair in 3 M KCl solution for about 4 h, and the output current increased to about 600 nA (supporting information).

In this work, when gaphene or RGO films were introduced to the experiments, no electrodes were exposed to the solution, as shown in Fig.3a ( the electrodes were covered with clear paste and were far away from the solution ). Thus, it is impossible for the electrodes to react with the solutions. This further demonstrated that the electricity was generated in this experiment and the experiment with one piece of graphene was not resulted from the chemical reaction between the electrodes and the solution.

A possible dynamic fluctuation mechanism might be used to explain the experiment results. It is well known that there is an electrical double layer on the surface of the pencil leads, RGO film or grahene film. The stern layer of the electrical double layer might be disturbed continually by the thermal motion of the ions within the double layer and lead to dynamic fluctuation to the charge transfer processes. But there is still no detail about the mechanism now.

In summary, we present an interesting idea of generating energy from the thermal motion of ions in the aqueous electrolyte solution by pencil leads, reduced graphene oxide films and graphene films. But the exact mechanism is still unclear. The results show that the thermal energy at room temperature can be harvested by low dimensional materials, for example graphene, or by the surface of solid material in the absence of temperature gradient. This finding will strongly promote the research of the self-powered technology.

**Acknowledgment:**

We thank Miss. L. Zhou, Ms. C. Qin ,Dr. T. H. Li, Mr. Z. Z. Zhang , Prof. H. X. Chang, and Prof. K. H. Wong for the fruitful discussion.

# Supporting Information:

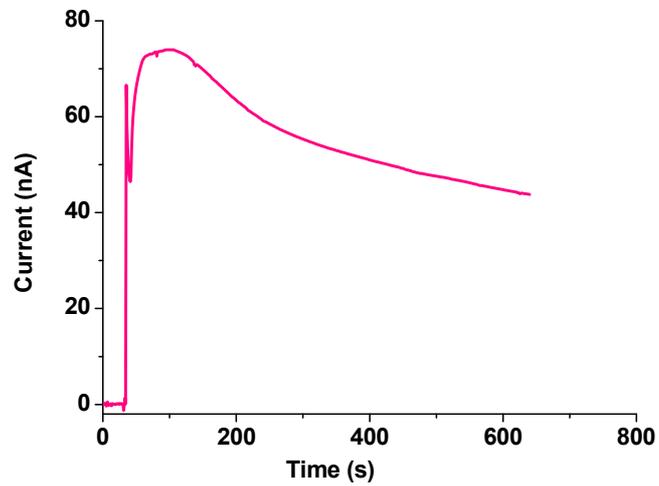

**Fig.S1.** Output of big HB-small HB pair.

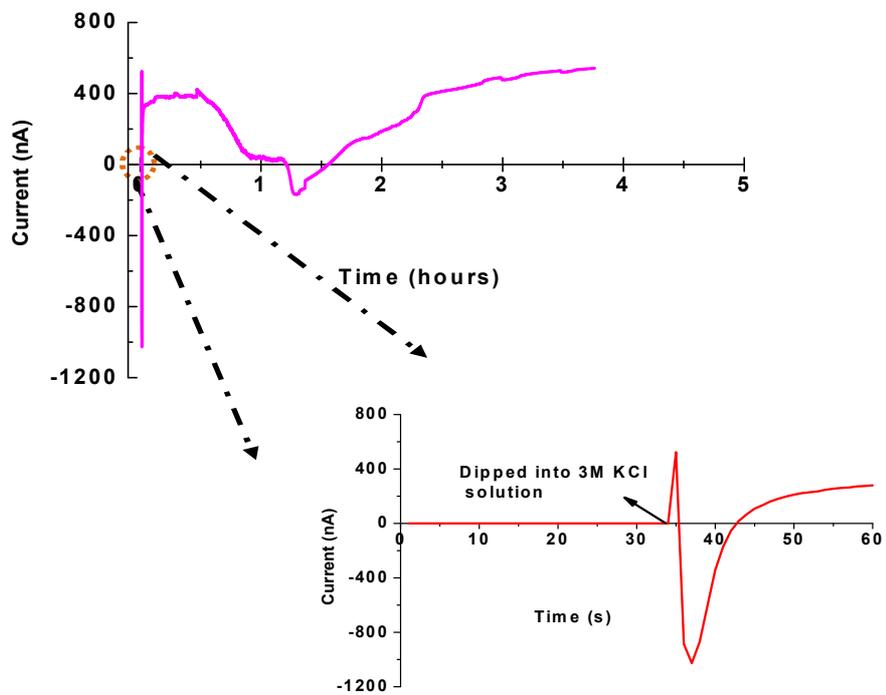

**Fig. S2.** Output of graphene-graphene pair in 3 M KCl solution.